\documentclass[preprint,12pt]{elsarticle}

\usepackage[margin=3cm]{geometry}
\usepackage{amsmath,amsfonts,amsthm}
\usepackage{algorithm}
\usepackage{algpseudocode}
\usepackage{array}
\usepackage[caption=false,font=normalsize,labelfont=sf,textfont=sf]{subfig}
\usepackage{textcomp}
\usepackage{stfloats}
\usepackage{url}
\usepackage{verbatim}
\usepackage{graphicx}
\usepackage{enumitem}
\usepackage{tikz}

\usepackage{multirow, multicol}
\usetikzlibrary{positioning, shapes.geometric, arrows.meta, fit, calc}

\tikzset{
	block/.style = {draw, rectangle, rounded corners, align=center, font=\small, minimum height=1.2em, minimum width=4.5cm},
	splitblock/.style = {draw, rectangle, rounded corners, align=center, font=\small, minimum height=2.5em, minimum width=5cm},
	wideblock/.style = {draw, rectangle, rounded corners, align=center, font=\small, minimum height=2.5em, minimum width=8cm}, 
	arrow/.style = {thick, -{Latex[width=1.5mm,length=1.5mm]}}
}

\tikzstyle{block} = [rectangle, draw, rounded corners, text centered, text width=3.6cm, minimum height=1.4cm, fill=blue!5]
\tikzstyle{arrow} = [-{Latex[length=2mm]}, thick]

\algrenewcommand\algorithmicrequire{\textbf{Input:}}
\algrenewcommand\algorithmicensure{\textbf{Output:}}

\newtheorem{remark}{Remark}
\newtheorem{theorem}{Theorem}
\newtheorem{definition}{Definition}
\newtheorem{lemma}{Lemma}

\def\BibTeX{{\rm B\kern-.05em{\sc i\kern-.025em b}\kern-.08em
		T\kern-.1667em\lower.7ex\hbox{E}\kern-.125emX}}
\usepackage{balance}

\begin{document}

\begin{frontmatter}



\title{A Novel Post-Quantum Secure Digital Signature Scheme Based on Neural Network} 


\author[label1]{Satish Kumar$^*$}

\author[label2]{Md. Arzoo Jamal} 

\affiliation[label1]{organization={Department of Computer Science and Engineering, Shiv Nadar University},
            city={Chennai},
            postcode={603110}, 
            state={Tamil Nadu},
            country={India}}


\affiliation[label2]{organization={Department of Aerospace Engineering, Indian Institute of Technology},
	city={Kanpur},
	postcode={208016}, 
	            state={Uttar Pradesh},
	country={India}}

\cortext[label1]{Satish Kumar. Email Id: satishkumar@snuchennai.edu.in}
\begin{abstract}
	Digital signatures are fundamental cryptographic primitives that ensure the authenticity and integrity of digital documents. In the post-quantum era, classical public key-based signature schemes become vulnerable to brute-force and key-recovery attacks due to the computational power of quantum algorithms. Multivariate polynomial based signature schemes are among the one of the cryptographic constructions that offers strong security guarantees against such quantum threats. With the growing capabilities of neural networks, it is natural to explore their potential application in the design of cryptographic primitives. Neural networks inherently captures the non-linear relationships within the data, which are encoded in their synaptic weight matrices and bias vectors. In this paper, we propose a novel construction of a multivariate polynomial based digital signature scheme that leverages neural network architectures. A neural network with binary weights is employed to define the central structure of the signature scheme. The design introduces a recurrent random vector, functionally analogous to an attention mechanism,  which contributes dynamic randomness based on the previous state, thereby enhancing the scheme's security. It is  demonstrated that the proposed signature scheme  provide security against Existential Unforgeability under adaptive Chosen-Message Attacks (EUF-CMA). Furthermore, it is proven that direct attacks aimed to recover the private keys are computationally infeasible within polynomial time, even in the presence of quantum computing abilities. The operational characteristics of the proposed scheme are also evaluated, with results indicating notable efficiency and practical viability in post-quantum cryptographic applications.
\end{abstract}




\end{frontmatter}



\section{Introduction}
In contemporary communication systems, ensuring data privacy, authenticity, and integrity has become a critical concern. Cryptography addresses these challenges by providing mechanisms to secure information against unauthorized access and tampering. In 1998, P. Shor introduced a polynomial-time algorithm capable of solving the Discrete Logarithm Problem using quantum computers, thereby enabling efficient key recovery attacks \cite{4}. Subsequently, researchers demonstrated that quantum computers could also solve the integer factorization problem in polynomial time. These developments highlighted the vulnerability of many classical cryptographic schemes to quantum attacks. As a result, the global cryptographic community has focused on developing new cryptographic algorithms that remain secure even in the presence of quantum adversaries. This shift has given rise to the field of post-quantum cryptography. Among the various approaches within this domain, multivariate polynomial-based public key cryptography has emerged as a prominent candidate for achieving quantum-resistant security. 
Although multivariate polynomial-based cryptographic primitives are generally considered secure against direct attacks, they suffer from practical inefficiency due to their large key sizes. Various modifications have been proposed to address this limitation, but with limited practical impact. In this paper, we propose a novel approach that leverages the power of matrix-based techniques to significantly reduce key size while maintaining the same security.

\subsection{Multivariate Polynomial Based Public Key Cryptography}
The security of multivariate polynomial-based public key cryptosystems relies on the computational difficulty of solving the systems of multivariate polynomial equations over the finite field. When the polynomials involved are of degree 2, the problem is referred to as the Multivariate Quadratic (MQ) problem, which is known to be NP-complete. Furthermore, systems involving higher-degree multivariate polynomials can typically be reduced to quadratic form through straightforward algebraic transformations. Consequently, the security of multivariate polynomial-based cryptosystems, even when utilizing polynomials of degree greater than 2, is effectively equivalent to the hardness of the MQ problem.

Over the years, various multivariate polynomial-based public key cryptosystems have been proposed. In 1985, Matsumoto and Imai \cite{10} introduced the first such system, known as the MI (Matsumoto-Imai) cryptosystem. However, it was later shown by Patarin \cite{15} that the MI cryptosystem is vulnerable to Linearization attacks, rendering it insecure. Building upon insights from this vulnerability, Patarin subsequently proposed a new and widely studied digital signature scheme known as the Oil-Vinegar (OV) Signature Scheme \cite{15}. Different versions of OV signature schemes have been proposed depending upon the security analysis, for more detail one can refer \cite{6}. In 1996, Patarin \cite{11} proposed a new cryptosystem named as Hidden Field Equation (HFE) based cryptosystem and it can be viewed as the extension of MI cryptosystem. Later different variants of HFE cryptosystems have been proposed \cite{6}. 
All of the discussed cryptosystems was using the properties of multivariate polynomials over the finite field.

In 2008, Gligoroski et al. \cite{16} introduced a multivariate polynomial-based cryptosystem whose central trapdoor function was constructed using Multivariate Quadratic Quasigroups (MQQ). However, subsequent analysis revealed that the scheme was vulnerable to various cryptanalytic techniques, including Gr{\"o}bner basis attacks, the MutantXL attack, and others \cite{12}. In response, Gligoroski et al. revised the original MQQ structure in 2012 and proposed two improved schemes: a signature scheme, MQQ-SIG \cite{13}, and an encryption scheme, MQQ-ENC \cite{14}. Nevertheless, in 2015, Faug{\'e}re et al. demonstrated that both schemes were susceptible to key recovery attacks, showing that it is possible to compute an equivalent key in polynomial time. In 2025, S. Kumar et al. \cite{8} proposed a new signature scheme named MQQ-SIGv, and provided evidence that, in contrast to earlier versions, recovering an equivalent key for this scheme is not feasible in polynomial time. For more information about public key cryptosystems using quasigroups, one can refer \cite{7}.

\subsection{Cryptosystems Based on Neural Networks}
The rise in the application of neural networks naturally raises the 
question of whether neural network architectures can be used to design cryptographic primitives. Several attempts have been made to develop such primitives. For different key exchange protocols using tree parity based neural networks, one can refer \cite{9}. Generative Adversarial network (GAN) based  cryptosystems discussed in \cite{17}.

In \cite{3}, C. Chan and L. M. Cheng analyzed the convergence properties of the clipped Hopfield neural network and used this framework to design a Linear Feedback Shift Register (LFSR) keystream generator. Building on this work, J. Wang, L. M. Cheng, and T. Su proposed a multivariate cryptographic scheme based on an extended clipped Hopfield neural network \cite{1}. Their proposed encryption method relies on the Discrete Logarithm Problem over matrices as its underlying hard problem. It has been demonstrated that this scheme can resist quantum attacks aimed at key recovery within polynomial time. However, in \cite{2}, S. Dai showed that quantum cryptanalysis techniques could recover the key in polynomial time, thus challenging the claimed quantum resistance.

Motivated by the aforementioned research, we design a multivariate polynomial-based signature scheme utilizing neural networks. In this paper, we present the following contributions along with the organization of the paper:

\begin{itemize}
	\item In Section \ref{sec:NN}, a modified Hopfield Neural Network incorporating binary weights and a recurrent attention vector is defined. Subsequently in Section \ref{sec:nntomvc}, this neural network is transformed into a system of multivariate polynomials.
	
	\item In Section \ref{sec:dlpmdp}, a computationally hard problem is formulated based on the  synaptic weight matrix and bias vector derived from the neural network. This problem remains difficult to solve even with quantum computing capabilities, and its complexity is analyzed.
	
	\item In Section \ref{sec:sig},  a digital signature scheme based on the proposed computational hard problem and the neural network framework is proposed. First  the bias vector is synchronized  between both users, after which the signature scheme comprising a tuple of algorithms (KeyGen, Sign, Verify) is described.
	
	\item In Section \ref{sec:securityanalysis}, we present a thorough security analysis of the signature scheme, demonstrating its resistance to direct attacks and proving its Existential Unforgeability under adaptive Chosen Message Attacks (EUF-CMA). Additionally, in Section \ref{sec:operationchar}, the operational characteristics of the signature scheme are defined and compare with  those of other proposed multivariate polynomial based signature schemes. Finally in Section \ref{sec:conclusion}, conclusions have been drawn for the paper.
\end{itemize}

\section{Binary Weight Networks (BWNs) with Recurrent Random Vector}\label{sec:NN}
A Hopfield neural network \cite{18}, also known as an associative memory, is a form of recurrent neural network designed to function as a content-addressable memory system. It comprises a single layer of neurons in which each neuron is interconnected with every other neuron, excluding self-connections. These interconnections are both bidirectional and symmetric, implying that the weight of the connection from neuron $i$ to neuron $j$ is equal to the weight from neuron $j$ to neuron $i$. The network recalls stored patterns through associative memory by fixing certain input values and allowing the system to evolve dynamically. This evolution is guided by the minimization of an energy function, ultimately converging to local energy minima that correspond to the stored memory patterns.

We propose the construction of  a network analogous to the Hopfield network, wherein each artificial neuron receives $n$ inputs, each associated with a synaptic weight.
The output of a neuron is determined by the aggregate sum of its  weighted inputs combined with a random component. Let the state of $i^{th}$ neuron at a time step $t$ be denoted by $S_{i,t}$.  The subsequent state $S_{i,t+1}$ is a function of the current states of all the other neurons and the corresponding weights, and is updated according to the following relation: 
\begin{equation}\label{eq:statefunc}
	S_{i,t+1} = f\left(\sum_{j=1}^{n}W_{i,j}\cdot(A_{j,t}\odot S_{j,t})+\theta_i\right),~ i=1,\dots,n.
\end{equation}
where $\odot$ denotes the element wise multiplication, $W_{i,j}$ represents the synaptic weight between neuron $i$ and $j$, $\theta_i$ denotes the threshold value of the neuron $i$ (also, known as bias corresponding to neuron $i$). The function $f(\cdot)$ is a nonlinear activation function that determines the neuron's output. Value of $A_{j,t}$ is the sigmoid based attention incorporating both the randomness and previous state $S_{j,t-1}$, can be calculated as:
\begin{equation}
	A_{j,t} = \sigma(S_{j,t-1}+\epsilon_{j,t}),
\end{equation} where $\epsilon_{j,t}\in U[0,1]$ represents the uniform distribution denote the randomness and $\sigma(z) = \frac{1}{1+e^{-z}}$ represents the Sigmoid function.
Generally, each neuron has two state one is firing state and other is quiescent state (can be represented as $S_{i,t}=1$ and $S_{i,t}=0$ respectively).

The synaptic weights $W_{i,j}^{Real}$ in the network may assume any real value, which poses challenges for practical hardware implementation due to increased computational and storage complexity. To address this issue, we employ binary weights, can be derived from the real valued weights using the Signum function, defined as follows:
\begin{equation}
	W_{i,j} = sgn(W_{i,j}^{Real}) = 
	\begin{cases}
		+1, ~\text{if} ~W_{i,j}^{Real}\ge 0\\
		-1, ~\text{if} ~W_{i,j}^{Real}< 0
	\end{cases}
\end{equation}
Where $sgn(\cdot)$ denotes the Signum function. The binarization significantly simplifies the implementation while preserving the essential polarity of the synaptic connections. In \cite{3}, it has been shown that clipped Hopfield neural network (CHNN) exhibits the convergence properties that make it suitable for applications such as the design of the keystream generators. Furthermore, it has been theoretically proved that the CHNN exhibits NP-complete characteristics, highlighting the inherent computational complexity associated with its dynamics and optimization behavior.  In this paper, the modified neural network can been interpreted  as an extended version of CHNN.  For the network defined by Equation (\ref{eq:statefunc}), the  nonlinear function $f(\cdot)$ is defined as:
\begin{equation}\label{eq:modulofunc}
	f(x)=
	\begin{cases}
		x\mod p, ~&\text{if}~ x \ge 0\\
		p-|x| \mod p ~&\text{if}~ x < 0
	\end{cases}
\end{equation}

where $p$ is a prime number. This design implies that each neuron can encode $(\lfloor\log_2p\rfloor+1)$ bits of information, $\lfloor\cdot\rfloor$ represents the floor function. Equation (\ref{eq:modulofunc}) plays a significant role in introducing the non-linearity into the multivariate polynomials, which make it suitable to design the quantum secure signature scheme.

If we fix $A_{j,t} = 1$ in Equation (\ref{eq:statefunc}), the resulting updated state becomes $S_{i,t+1} = f\left(\sum_{j=1}^{n}W_{i,j}\cdot S_{j,t}+\theta_i\right)$, $ i=1,\dots,n$, which is identical to the update state of the Clipped Hopfield neural network as defined in \cite{3}. This implies that the
update state described by Equation (\ref{eq:statefunc}), in the proposed network can be interpreted as a generalized form of Clipped Hopfield neural network introduced in \cite{3}.

\section{Conversion of Binary Weight Neural Network to Non-Linear Multivariate Polynomials}\label{sec:nntomvc}
We will map the proposed neural network defined by Equation (\ref{eq:statefunc}) to the multivariate polynomial. 
It can be observed that Equation (\ref{eq:statefunc}), the state $S_{i,t+1}$ can be expressed in matrix form. As $t$ increases, this equation may be interpreted as an enhanced multivariate polynomials that employs a large set of low degree modular equations defined over a large number modulus. The resulting system of equations can be solved using both iterative and polynomial methods. Thus, the network defined by Equation (\ref{eq:statefunc}) can be effectively mapped onto multivariate polynomials and for $i=1,\dots,n$, the equation could be written as: 
\begin{align}\label{eq:systemofe}
	S_{1,t+1} &= f\left(\sum_{j=1}^{n}W_{1,j}(A_{j,t}\odot S_{j,t})+\theta_1\right)\nonumber\\
	S_{2,t+1} &= f\left(\sum_{j=1}^{n}W_{2,j}(A_{j,t}\odot S_{j,t})+\theta_2\right)\nonumber\\
	&\vdots\nonumber\\
	S_{n,t+1} &= f\left(\sum_{j=1}^{n}W_{n,j}(A_{j,t}\odot S_{j,t})+\theta_n\right).
\end{align}
Conclusively, system of Equations (\ref{eq:systemofe}) can be  reformulated and expressed as:
\begin{equation}\label{eq:1}
	S_{t+1} = f(W\cdot(A_t\odot S_t) +\theta)
\end{equation}
where 
$
W = 
\begin{bmatrix}
	W_{1,1} & W_{1,2} & \cdots & W_{1,n} \\
	W_{2,1} & W_{2,2} & \cdots & W_{2,n} \\
	\vdots  & \vdots & \ddots & \vdots \\
	W_{n,1} & W_{n,2} & \cdots & W_{n,n}
\end{bmatrix},
S_t =\begin{bmatrix}
	S_{1,t}\\
	S_{2,t}\\
	\vdots\\
	S_{n,t}
\end{bmatrix},~
A_{t}=
\begin{bmatrix}
	A_{1,t}\\
	A_{2,t}\\
	\vdots\\
	A_{n,t}
\end{bmatrix}
~\text{and}~
\theta=
\begin{bmatrix}
	\theta_{1}\\
	\theta_{2}\\
	\vdots\\
	\theta_{n}
\end{bmatrix} 
$.
Equation (\ref{eq:1}) shows that the network system can be constructed iteratively, wherein the subsequent neuronal state $S_{i,t+1}$ is derived from the current state $S_t$. From an implementation standpoint, the synaptic weight matrix is constrained to binary values, specifically $W_{i,j}\in\{-1,1\}$, and it is imperative that the  matrix $W$ possess non-zero determinant, to ensure that $W^{-1}$ exists. This requirement is critical for the decryption or signature operations of documents on hardware with limited numerical precision.

Let initial state of the network at $t=0$ is denoted by $S_0$, and assume that the $f(\cdot)$ function corresponds to a  modulo operation as defined in Equation (\ref{eq:modulofunc}), adhering to the  properties inherent to modulo arithmetic.
The recurrence relation incorporating random attention at each time step is given by:
\begin{equation}\label{eq:At}
	S_{t+1} = f(W\cdot(A_t\odot S_t)+\theta) = f(\widetilde{W}_tS_t+\theta),
\end{equation}
where $\widetilde{W}_j = W\cdot diag(A_j)$. Unfolding this recurrence  to $\rho$ steps yields the following expression:
\begin{equation}
	S_\rho = f\left(\widetilde{W}_{\rho-1}\cdot f(\widetilde{W}_{\rho-2}\dots f(\widetilde{W}_0\cdot S_0+\theta)\dots +\theta)+\theta \right).
\end{equation}
Above recurrence relation  can be approximated and expressed as: 
\begin{equation}\label{eq:recurrencerelation}
	S_\rho \approx f\left(\underbrace{\left(\prod_{j=0}^{\rho-1}\widetilde{W}_j\right)\cdot S_0}_\text{Transformed initial state}+\underbrace{\sum_{k=0}^{\rho-1}\left(\prod_{j=k+1}^{\rho-1}\widetilde{W}_j\right)\cdot\theta}_\text{Accumulated bias terms}\right),
\end{equation}


where $\widetilde{W_j} = W\cdot diag(A_j)$, does not commute with $W$ and $\prod_{j=0}^n\widetilde{W}_j \neq W^n\cdot diag(\prod A_j)$. Note that $A_j$ is an $n$-dimensional vector and $diag(A_j)$ is a $n\times n$ matrix where diagonal elements can be represented by corresponding element of vector $A_j$ and other elements of matrix is zero. Additionally, $\prod_{j=0}^{\rho-1}\widetilde{W}_j=\widetilde{W}_{\rho-1}\cdot\widetilde{W}_{\rho-2}\dots\widetilde{W}_0$ (right-to-left multiplication).
Eventually, the neural network converges to the state $S_{\rho} = Y$, where $Y$ constitutes the solution to the corresponding system of multivariate polynomial equations
$
\left(\prod_{j=0}^{\rho-1}\widetilde{W}_j\right)\cdot S_0+{\sum_{k=0}^{\rho-1}\left(\prod_{j=k+1}^{\rho-1}\widetilde{W}_j\right)\cdot\theta}
$ 
with respect to input initial vector, it can be expressed as: 
$$
X=
\begin{bmatrix}
	x_1&x_2 & \cdots & x_n
\end{bmatrix}^T = S_0 =
\begin{bmatrix}
	S_{1,0} & S_{2,0} & \cdots & S_{n,0}
\end{bmatrix}^T
$$
Equation (\ref{eq:recurrencerelation}) indicates that the state of the neural network is determined by the initial state $S_0$, the iteration number $\rho$, and the threshold vector $\theta$. This relationship can be reformulated as a system of multivariate polynomials form in field $\mathbb{Z}_p$ as follows:
\begin{equation*}
	Y = f\left(\left(\prod_{j=0}^{\rho-1}\widetilde{W}_j\right)\cdot S_0+{\sum_{k=0}^{\rho-1}\left(\prod_{j=k+1}^{\rho-1}\widetilde{W}_j\right)\cdot\theta}\right)
\end{equation*} and rewritten in the form of 
\begin{equation}\label{eq:mvcpoly}
	Y = f(W_xX+W_{\theta}\theta)
\end{equation}
where $W_x = \left(\prod_{j=0}^{\rho-1}\widetilde{W}_j\right)$ and $W_{\theta}=\sum_{k=0}^{\rho-1}\left(\prod_{j=k+1}^{\rho-1}\widetilde{W}_j\right)$.
From Equation (\ref{eq:mvcpoly}), the initial state of the network, namely the input vector $X$, can be readily recovered using:
\begin{align}\label{eq:mvcinvpoly}
	X = f(W_x^{-1}(Y-f(W_{\theta}\theta))).
\end{align}

At this stage, the correspondence between the defined neural network architecture and a multivariate polynomial based framework formally established. This mapping is particularly significant in the context of complexity theory and cryptographic applications. The neural networks output exhibits intricate, nonlinear dynamics that are effectively captured by the multivariate formulation. In Equation (\ref{eq:At}), the vector $A_t$
directly obscure the relationship between the message and it corresponding signature, which making it difficult for an attacker to deduce the original synaptic weights or (message,signature) pair even it has access to the neural network architecture, famously called Confusion as per Shanon Theory \cite{19}.
Most importantly, this transformation preserves the NP-hard nature and non-linearity of the original problem, thereby maintaining the computational intractability that underpins its cryptographic strength.  

\section{Discrete Logarithm with Matrix Decomposition Problem (DL-MDP) }\label{sec:dlpmdp}
In this section, we formally give the notion of  Combined Discrete Logarithm and Matrix Decomposition Problem (DL-MDP), and provide an analysis of its brute-force complexity as well as its security implications in both classical and quantum computational settings.

\begin{definition}[Discrete Logarithm Problem (DLP) \cite{19}]
	For given prime $p$, a generator $\alpha\in\mathbb{Z}_p^*$ and an element $\beta\in\mathbb{Z}_p^*$, where $\mathbb{Z}_p^*$ is a cyclic group. Find an integer $x, 0\le x\le p-2$ such that $\alpha^x = \beta\pmod p$.
\end{definition}

\begin{definition}[Matrix Decomposition Problem (MDP) \cite{19}]
	For given matrix $M = \mathbb{Z}_p^{m\times n}$, the matrix decomposition problem asks to find two matrices $A\in \mathbb{Z}_p^{m\times k}$ and $B\in \mathbb{Z}_p^{k\times n}$ such that: $M = A\cdot B$.
\end{definition}

In \cite{2}, S. Dai demonstrated that the Discrete Logarithm Problem (DLP) for matrices can be efficiently solved using quantum computing in polynomial time, posing a significant threat to multivariate cryptosystems that rely on the hardness of this problem, such as the one proposed in \cite{1}. To address this vulnerability, we propose a hybrid approach that combines the Discrete Logarithm (DLP) with the Matrix Decomposition Problem (MDP). Specifically, we raise the weight matrix to a power and then increase complexity by permuting the rows of the resulting matrix through multiplication with a permutation matrix. This added layer of transformation enhances the overall brute force complexity of the problem.

\begin{definition}
	Let $GL(n,\mathbb{Z}_p)$ be a general linear group over $\mathbb{Z}_p$ and $\mathbb{Z}_p^*$ be a finite cyclic group. Let $A,B,L$ be an arbitrary element of $GL(n,\mathbb{Z}_p)$ and $a\in\mathbb{Z}_p^*$. Then, for given $A,B\in GL(n,\mathbb{Z}_p)$ such that $B = L\cdot A^a$, find the tuple $(L,a)$, where  $L\in GL(n,\mathbb{Z}_p)$ and $a\in\mathbb{Z}_p^*$. This is named as ``Discrete Logarithm with Matrix Decomposition Problem (DL-MDP)''.
\end{definition}

In the above definition if either $a$ or  $L$ is given,  the problem DL-MDP reduces to either DLP or MDP,  as follows:
\begin{itemize}
	\item If $a\in\mathbb{Z}_p^*$ is given, the equation $B = L\cdot A^a$ reduces to $B = L\cdot A'$, where $A'=A^a$. The problem now becomes, for given $B$ and $A'$, to find the tuple $L$ such that the equation $B = L\cdot A'$ holds, which is exactly the Matrix Decomposition Problem (MDP).
	
	\item If the tuple $L$ is given, the equation $B = L\cdot A^a$ becomes $L^{-1}\cdot B= A^a$, which can be rewritten as $B' = A^a$, where $B'=L^{-1}\cdot B$.  The problem now is now to find $a\in\mathbb{Z}_p^*$ such that $B' = A^a$, which is the Discrete Log Problem (DLP) problem.
\end{itemize}

Thus we can conclude that the DL-MDP is a combination of the DLP and MDP. The complexity of the DL-MDP problem and the level of security of cryptosystems based on the DL-MDP  depend on the order of $L$ as $n$ and prime $p$.

\begin{theorem}
	Let $A,B\in GL(n,\mathbb{Z}_p)$, where $\mathbb{Z}_p$ is a finite field with prime order $p$ and $L\in GL(n,\mathbb{Z}_p)$ are permutation matrices. The problem of finding a tuple $(a,L)$, where $a\in\mathbb{Z}_p$, such that:
	\begin{equation}
		B = L\cdot A^a \pmod p
	\end{equation}
	can be solved using a brute force approach with time complexity $\mathcal{O}((p-1).n!.n^3\log p)$.
	
	\begin{proof}
		The value of $a$ can take any of the possible $p-1$ values in $\mathbb{Z}_p^*$ and $L$ are permutation matrices, which yield $n!$ combinations. Consequently, computing each tuple $(a,L)$ such that for given $A,B\in GL(n,\mathbb{Z}_p)$ such that $B = L\cdot A^a \pmod p$  requires $\mathcal{O}(n^3\log p)$ steps.  So, the total brute force complexity is $\mathcal{O}((p-1).n!.n^3\log p)$. 
	\end{proof}
\end{theorem}

\begin{remark}
	To achieve greater than 128-bit security against brute force attacks using classical computers, the parameter pair $(n,p)$ must satisfy the inequality\\ $\log_2\left((p-1).n!.n^3\log p\right)\ge {128}$.
\end{remark}

\begin{remark}
	When considering quantum attacks using Grover's Algorithm \cite{5}, achieving 128-bit security or more against brute force attacks requires selecting the parameter  pair $(n,p)$ such that the inequality $\log_2\left((p-1).n!.n^3\log p\right)\ge{256}$ holds.
\end{remark}


\begin{remark}
	
	Without loss of generality, achieving $\ell$-bit security against brute-force attacks requires choosing parameters $(n, p)$ such that 
	\begin{equation}
		\log_2\left((p - 1) \cdot n! \cdot n^3 \log p\right) \ge \ell
	\end{equation}
	for classical adversaries. In the case of quantum adversaries, the requirement becomes
	\begin{equation}
		2 \log_2\left((p - 1) \cdot n! \cdot n^3 \log p\right) \ge \ell.
	\end{equation}
\end{remark}

\section{Signature Scheme}\label{sec:sig}
In this section, we propose a signature scheme whose security is based on the hardness of the DL-MDP problem, defined in Section \ref{sec:dlpmdp}. The scheme consists of a tuple of three algorithms: (KeyGen, Sign, verify), where KeyGen algorithm generates the public and private key pair, Sign algorithm produces the signature for the given message of document, and Verify algorithm checks the authenticity of a given signature for given signature and message pair.
Suppose user A wants to sign a document or message $M\in\mathbb{Z}_p^n$ and user B wants to verify it. Prior to any communication, both users must initialize their environment. This includes setting up their respective networks and agreeing upon the synaptic weights $W$, a prime number $p$, an integer $n$, and a randomly chosen vector $Q$.

\subsection{Setup the Environment for Signature}
Suppose user A and user B wants to communicate with each other or want to the authenticity of the documents using digital signatures. Both users   runs the neural network defined using the Equation (\ref{eq:statefunc}) once and  agree upon the synaptic weights $W$ (i.e., both users must have same weights), vector $A_j$ for fixed $j=0,1,\dots,\rho$, which contains the essence of randomness (Note that step $\rho$ will be kept secret between both users), prime $p$, integer $n$ and a random vector $Q\in\mathbb{Z}_p^n$. Additionally, they must synchronize the bias vector prior to any communication. This synchronization process can be performed using the protocol outlined in Table \ref{tab:synvector}.

\subsection{Threshold Vector Synchronization}
Let $\mathbb{Z}_p$ denote a finite field, where $p$ is a prime number. To enable vector synchronization between user A and user B, a mask vector is first generated, whose security relies on the computational hardness of the discrete log problem over matrices. Utilizing there respective public keys along with the masked vector, both users are able to synchronize their internal vectors. 

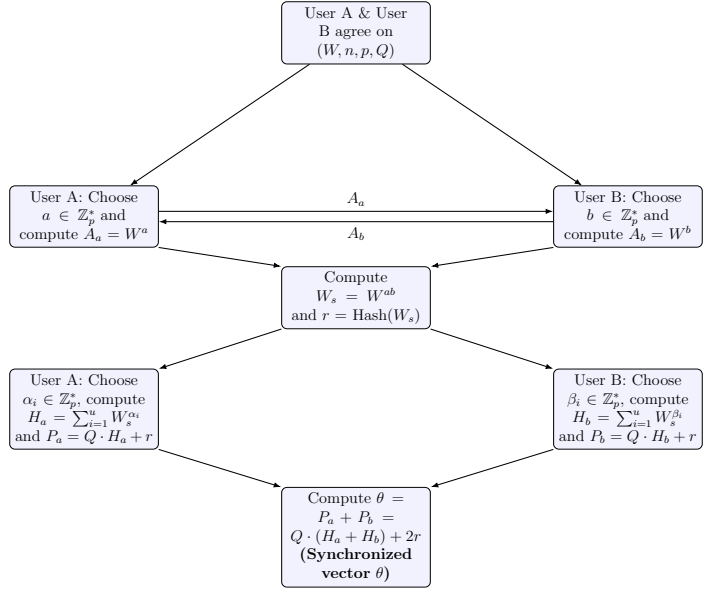
\begin{figure}[]
	\centering
	\resizebox{0.6\textwidth}{0.5\textwidth}
	{
		\begin{tikzpicture}[node distance=2.2cm and 1.5cm]  
			
			\node[block] (agree) {User A \& User B agree on $(W, n, p, Q)$};
			
			\node[block, below left=3.3cm and 3.2cm of agree] (a1) {User A: Choose $a \in \mathbb{Z}_p^*$ and compute $A_a = W^a$};
			
			\node[block, below right=3.3cm and 3.2cm of agree] (b1) {User B: Choose $b \in \mathbb{Z}_p^*$ and compute $A_b = W^b$};
			
			\node[block, below=5.5cm of agree] (shared) {Compute $W_s = W^{ab}$ and $r = \text{Hash}(W_s)$};
			
			\node[block, below=3.3cm of a1] (a2) {User A: Choose $\alpha_i \in \mathbb{Z}_p^*$, compute \\ $H_a = \sum_{i=1}^{u} W_s^{\alpha_i}$ \\ and $P_a = Q \cdot H_a + r$};
			
			\node[block, below=3.3cm of b1] (b2) {User B: Choose $\beta_i \in \mathbb{Z}_p^*$, compute \\ $H_b = \sum_{i=1}^{u} W_s^{\beta_i}$ \\ and $P_b = Q \cdot H_b + r$};
			
			\node[block, below=4.3cm of shared] (theta) {Compute $\theta = P_a + P_b = Q \cdot (H_a + H_b) + 2r$ \\ \textbf{(Synchronized vector $\theta$)}};
			
			\draw[arrow] (agree) -- (a1);
			\draw[arrow] (agree) -- (b1);
			
			\draw[arrow] (a1.south east) -- (shared.north west);
			\draw[arrow] (b1.south west) -- (shared.north east);
			
			\draw[arrow] (shared.south west) -- (a2.north east);
			\draw[arrow] (shared.south east) -- (b2.north west);
			
			\draw[arrow] (a2.south east) -- (theta.north west);
			\draw[arrow] (b2.south west) -- (theta.north east);
			
			\draw[arrow] ([yshift=4pt]a1.east) -- ([yshift=4pt]b1.west) node[midway, above] {$A_a$};
			\draw[arrow] ([yshift=-4pt]b1.west) -- ([yshift=-4pt]a1.east) node[midway, below] {$A_b$};
			
		\end{tikzpicture}
	}
	\caption{Pictorial representation for synchronized threshold vector}
	\label{fig:thresholdvector}
\end{figure}

As discussed, suppose user A and user B runs the neural network, agrees on the weight matrix $W$, a randomly generated vector $Q = (q_1,\dots,q_n)$, where $q_i\in\mathbb{Z}_p, \forall i=1,\dots,n$ and a universal hash function $Hash:\mathbb{Z}_p^n\times \mathbb{Z}_p^n\rightarrow \mathbb{Z}_p^n$. To synchronize the vector both user A and user B follows protocol described in Table  \ref{tab:synvector}. For pictorial representation of the synchronization protocol, one can refer Figure \ref{fig:thresholdvector}.

\begin{table*}[t]
	\centering
	\begin{tabular}{|p{6.5cm}|p{6.5cm}|}
		\hline
		\multicolumn{2}{|p{13cm}|}{\centering\textbf{Generation of Masked Vector Based on Discrete Logarithm Problem}} \\
		\hline
		\textbf{User A} & \textbf{User B} \\
		\hline
		\begin{itemize}[leftmargin=*]
			\item User A picks random number $\texttt{a} \in \mathbb{Z}_p$.
			\item It computes $A_\texttt{a} = W^\texttt{a} \mod p$.
			\item Sends $A_\texttt{a}$ to User B.
		\end{itemize}
		&
		\begin{itemize}[leftmargin=*]
			\item User B picks random number $\texttt{b} \in \mathbb{Z}_p$.
			\item It computes $A_\texttt{b} = W^\texttt{b} \mod p$.
			\item Sends $A_\texttt{b}$ to User A.
		\end{itemize}
		\\
		\hline
		\multicolumn{2}{|p{13cm}|}{\centering
			The masked vector will be: $r = \text{Hash}(W_s) \in \mathbb{Z}_p^n$, where $W_s = W^{\texttt{a}\texttt{b}} \mod p$ is the shared matrix.
		} \\
		\hline
		\begin{enumerate}[leftmargin=*]
			\item For $i$ to $u$: Pick $\alpha_i \in \mathbb{Z}_p$.
			\item Compute $H_\texttt{a} = \sum_{i=1}^{u} W_s^{\alpha_i}$.
			\item Compute the public vector $P_\texttt{a} = Q \cdot H_\texttt{a} + r$.
		\end{enumerate}
		&
		\begin{enumerate}[leftmargin=*]
			\item For $i$ to $u$: Pick $\beta_i \in \mathbb{Z}_p$.
			
			\item Compute $H_\texttt{b} = \sum_{i=1}^{u} W_s^{\beta_i}$.
			\item Compute the public vector $P_\texttt{b} = Q \cdot H_\texttt{b} + r$.
		\end{enumerate}
		\\
		\hline
		\multicolumn{2}{|p{13cm}|}{\centering
			The shared synchronized vector will be $\theta = P_\texttt{a} + P_\texttt{b} = Q \cdot (H_\texttt{a} + H_\texttt{b}) + 2r$.
		} \\
		\hline
	\end{tabular}
	\caption{Synchronized vector generation protocol for users A and B}

	\label{tab:synvector}
\end{table*}

\subsection{Key Generation Protocol (KeyGen)}\label{sec:keygen}
In a digital signature scheme, the Key Generation (KeyGen) protocol generates a pair of cryptographic keys: A secret key and a corresponding public key that is made publicly available. Suppose user A intends to sign a message  $M\in\mathbb{Z}_p^n$. The signature is produced using 
Equation (\ref{eq:mvcinvpoly}) for given message.

Assume that user A possess a shared matrix $W$ and a set of vectors vector $A_j$ for time steps $j=0,\dots,\rho-1$. For each $j$, user A computes $W_j = W.diag(A_j)$. Using these matrices $W_j$, one can compute 

\begin{equation}\label{eq:wx}
	W_x = \prod_{j=0}^{\rho-1}W_j,\text{ and }  W_{\theta} = \sum_{k=0}^{\rho-1}(\prod_{j=k+1}^{\rho-1}W_j).
\end{equation}

Using these, the masked versions are computed as: 

\begin{align}
	\overline{W_x} &= L_x\cdot {W_x}^a= L_x\cdot(\prod_{j=0}^{\rho-1}W\cdot diag(A_j))^a \nonumber \\&= L_x\cdot(W\cdot(diag(A_1)\cdot diag(A_2)\cdot\dots \cdot diag(A_j))) ^a. 
\end{align}
and
\begin{equation}
	\overline{W_{\theta}} = L_{\theta}\cdot W_{\theta}^b = L_{\theta}\cdot (\sum_{k=0}^{\rho-1}(\prod_{j=k+1}^{\rho-1}W_j))^b.
\end{equation}

Here, the central matrices $W_x^a$ and $W_{\theta}^b$ are masked by the permutation matrices $L_x$ and $L_{\theta}$ respectively. 
The tuple $(L_x,L_{\theta}, a, b, \rho, W, A_j)$ is the private key tuple. The tuple $(\overline{W_x},\overline{W_{\theta}})$ is the public key tuple. 

\subsection{Signature Generation Protocol (Sign)}
We present a digital signature scheme for a document or message $M\in\mathbb{Z}_p^n$, where the construction is based on Equation (\ref{eq:mvcinvpoly}). The signing process utilizes a  private key tuple $(L_x,L_{\theta},a,b)$ in conjuction with the synaptic weight.

In the signing phase, the hash of the message $M$ is first concatenated   
with randomly selected elements from the field $\mathbb{Z}_p$. This concatenation serves to enhance the security of the scheme by increasing resistance to direct attacks. The resulting input string is then used, along with Equation (\ref{eq:mvcinvpoly}), to generate the corresponding signature.


\begin{algorithm}[]
	\caption{Signature of message $M$}
	\begin{algorithmic}[1]
		\Require Message $M=\{m_1,\dots,m_l,m_{l+1},\dots, m_n\}\in\mathbb{Z}_p^n$ and a secret key tuple  $(L_x,L_{\theta},a,b,\rho, W,A_j)$.
		\State Compute  $h = hash(m_1,\dots,m_l,m_{l+1},\dots,m_n) = h_0\mid\mid h_1$; here, $h_0$ is the first $l$ coordinates of $h$ and $h_1$ is remaining $n-l$ coordinates of $h$.
		
		\State Choose $r_0 = (r_{01},r_{02},\dots,r_{0(n-l)})\in\mathbb{Z}_P^{n-l}$ and $r_1 = (r_{11},r_{12},\dots,r_{1l})\in\mathbb{Z}_p^l$ randomly and uniformly. Subsequently, compute $x_0 = r_0\mid\mid h_0$ and $x_1 = r_1\mid\mid h_1$. Both $x_0$ and $x_1$ is $n$-bit long vector.
		
		\State Compute $\sigma_0 = f\left(W_x^{-a}L_x^{-1}(x_0-f((L_{\theta}.W_{\theta}^b)\cdot \theta))\right)$,
		and $\sigma_1 = f\left(W_x^{-a}L_x^{-1}(x_1-f((L_{\theta}.W_{\theta}^b)\cdot \theta))\right)$
		Here, $W_x$ and $W_{\theta}$ can be computed using Equation (\ref{eq:wx}).
		
		
		\Ensure The signature for corresponding to message $M\in\mathbb{Z}_p^n$ is $\sigma = (\sigma_0, \sigma_1)$.
	\end{algorithmic}
	\label{algo:sign}
\end{algorithm}

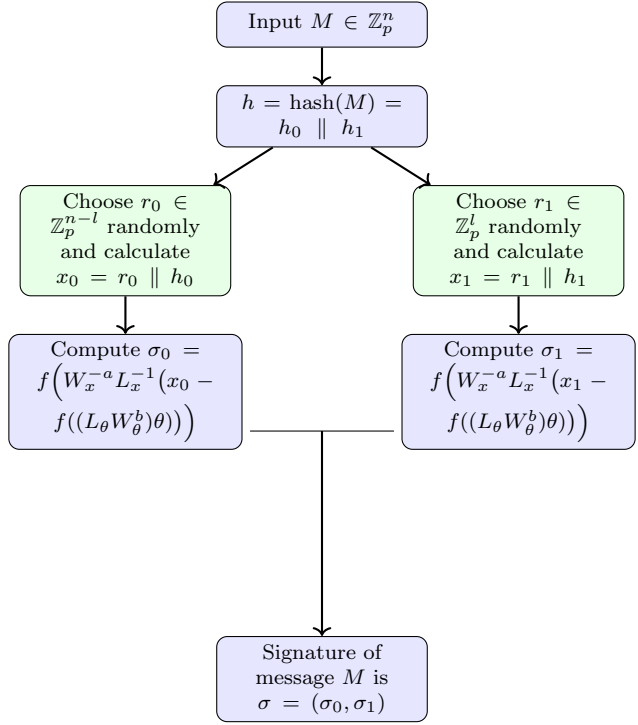
\begin{figure}[h]
	\centering
	\scriptsize
	\begin{tikzpicture}[
		node distance=0.5cm and 0.2cm,
		block/.style={rectangle, draw, fill=blue!10, text width=2.6cm, minimum height=0.6cm, text centered, rounded corners},
		splitblock/.style={rectangle, draw, fill=green!10, text width=2.6cm, minimum height=0.6cm, text centered, rounded corners},
		longblock/.style={rectangle, draw, fill=blue!10, text width=2.9cm, minimum height=1.0cm, text centered, rounded corners},
		arrow/.style={->, thick}
		]
		
		\node[block, xshift=0.3cm] (input) {Input $M \in \mathbb{Z}_p^n$};
		\node[block, below=of input] (hash) {$h = \text{hash}(M) = h_0 \parallel h_1$};
		
		\node[splitblock, below left=of hash, xshift=0.4cm] (choose0) {
			Choose $r_0 \in \mathbb{Z}_p^{n-l}$ randomly\\ and calculate $x_0 = r_0 \parallel h_0$
		};
		\node[splitblock, below right=of hash, xshift=-0.4cm] (choose1) {
			Choose $r_1 \in \mathbb{Z}_p^l$ randomly\\ and calculate $x_1 = r_1 \parallel h_1$
		};
		
		\node[longblock, below=of choose0] (sigma0) {
			Compute $\sigma_0 = f\Big(W_x^{-a}L_x^{-1} \big(x_0 - f((L_{\theta}W_{\theta}^b)\theta)\big)\Big)$
		};
		\node[longblock, below=of choose1] (sigma1) {
			Compute $\sigma_1 = f\Big(W_x^{-a}L_x^{-1} \big(x_1 - f((L_{\theta}W_{\theta}^b)\theta)\big)\Big)$
		};
		
		\node[block, below=1.2cm of hash, yshift=-5.3cm] (final) {
			Signature of message $M$ is $\sigma = (\sigma_0, \sigma_1)$
		};
		
		\draw[arrow] (input) -- (hash);
		\draw[arrow] (hash) -- (choose0);
		\draw[arrow] (hash) -- (choose1);
		\draw[arrow] (choose0) -- (sigma0);
		\draw[arrow] (choose1) -- (sigma1);
		
		\coordinate (start) at ($(sigma0.south east) + (0.1,0.25)$);
		\coordinate (end) at ($(sigma1.south west) + (-0.1,0.25)$);
		\draw[-] (start) -- (end);
		
		\path (start) -- (end) coordinate[midway] (mid);
		\draw[arrow] (mid) -- (final.north);
		
	\end{tikzpicture}
	\caption{Signature Generation Protocol}
	\label{fig:signature-generation}
\end{figure}

\subsection{Verification Protocol (Verify)}
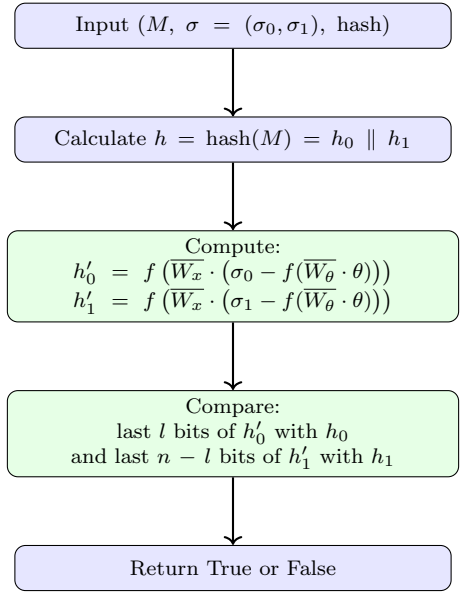
\begin{figure}[]
	\centering
	\scriptsize
	\begin{tikzpicture}[
		node distance=0.9cm and 0cm,
		block/.style={rectangle, draw, fill=blue!10, text width=5.6cm, minimum height=0.6cm, text centered, rounded corners},
		wideblock/.style={rectangle, draw, fill=green!10, text width=5.8cm, minimum height=0.8cm, text centered, rounded corners},
		arrow/.style={->, thick}
		]
		
		\node[block] (input) {Input $(M,\ \sigma = (\sigma_0, \sigma_1),\ \text{hash})$};
		
		\node[block, below=of input] (hash) {Calculate $h = \text{hash}(M) = h_0 \parallel h_1$};
		
		\node[wideblock, below=of hash] (compute) {
			Compute:\\
			$h_0' = f\left(\overline{W_x} \cdot \left(\sigma_0 - f(\overline{W_{\theta}} \cdot \theta)\right)\right)$\\
			$h_1' = f\left(\overline{W_x} \cdot \left(\sigma_1 - f(\overline{W_{\theta}} \cdot \theta)\right)\right)$
		};
		
		\node[wideblock, below=of compute] (compare) {
			Compare:\\
			last $l$ bits of $h_0'$ with $h_0$\\
			and last $n-l$ bits of $h_1'$ with $h_1$
		};
		
		\node[block, below=of compare] (result) {Return True or False};
		
		\draw[arrow] (input) -- (hash);
		\draw[arrow] (hash) -- (compute);
		\draw[arrow] (compute) -- (compare);
		\draw[arrow] (compare) -- (result);
		
	\end{tikzpicture}
	\caption{Verification Protocol}
	\label{fig:verification}
\end{figure}

In the verification protocol, the verifier is provided with  a message $M\in\mathbb{Z}_p^n$, a tuple $\sigma = (\sigma_0,\sigma_1)$, and a public key tuple $(\overline{W_x},\overline{W_\theta}, hash)$. The objective of the protocol is to determine the authenticity of the signature with respect to the given message. The verification is performed by executing the steps involved in  \ref{algo:verify}. For better understanding and pictorial representation of protocol, one can refer Figure \ref{fig:verification}.

\begin{algorithm}[H]
	\caption{Verification Protocol}
	\begin{algorithmic}[1]
		\Require A message $M\in\mathbb{Z}_p^n$, tuple $\sigma = (\sigma_0,\sigma_1)$ and public key pair $(\overline{W_x},\overline{W_\theta}, hash)$.
		
		\State Compute  $h = hash(m_1,\dots,m_l,m_{l+1},\dots,m_n) = h_0\mid\mid h_1$; here, $h_0$ is the first $l$ coordinates of $h$ and $h_1$ is remaining $n-l$ coordinates of $h$.
		
		
		\State Calculate $h_0' = f\left(\overline{W_x}\cdot(\sigma_0-f(\overline{W_{\theta}}\cdot\theta))\right)$ and $h_1' = f\left(\overline{W_x}\cdot(\sigma_1-f(W_\theta\cdot \theta))\right)$.
		
		\State Compare the last $l$ coordinates of $h_0'$ with $h_0$ and last $n-l$ coordinates of $h_1'$ with $h_1$. If each coordinates matches then return True else return False.

		\Ensure True or False.
	\end{algorithmic}
	\label{algo:verify}
\end{algorithm}

	%
	%
	%
	%
	%
	%
	%

\section{Security Analysis}\label{sec:securityanalysis}

In this section, we analyze the security of the proposed signature scheme and establish several important results.
\subsection{Existential Unforgeability under Chosen Message Attack (EUF-CMA) Security}

\begin{definition}
	A function $f:\mathbb{N}\mapsto \mathbb{R}^+$ is negligible if for every positive polynomial $p$ there is an $N$ such that for all integers $n>N$ it holds that $f(n)<\frac{1}{p(n)}$.
\end{definition}

Consider a digital signature scheme 
Sig =(KeyGen, Sign, Verify), where 
KeyGen denotes the key generation algorithm, 
Sign denotes the signing algorithm, and 
Verify denotes the verification algorithm. Assume the existence of an adversary $\mathcal{A}$
that interacts with the signing algorithm as a black-box oracle. Let $Ch$ denote the challenger in the security game. The experiment 
$Exp_{Sig(k)}^{EUF-CMA}$, modeling existential unforgeability under an adaptive chosen-message attack, is conducted between 
$\mathcal{A}$ and 
$Ch$ as described below:

\begin{algorithm}
	\caption{{\bf Experiment} $Exp_{Sig(k)}^{EUF-CMA}$}
	\begin{algorithmic}[1]
		\State The challenger $Ch$ generate key $(pk, sk)\leftarrow$ KeyGen$(k)$ by running the key generation algorithm which takes security parameter $k$ as an input. Additionally, give $pk$ to $\mathcal{A}$.
		
		\State The adversary $\mathcal{A}$ requests signatures for chosen $n$ messages $\{m_i\}_{i=1}^n$ and obtains the valid signatures $\{\sigma_i\}_{i=1}^n$ in response, where $\sigma_i\leftarrow \text(Sign)(sk, m_i)$ for $i=1,\dots, n$.
		
		\State The adversary $\mathcal{A}$ adaptively produces the message signature pair $(\sigma^*, m^*)$.
		
		\State The output of the experiment $Exp_{Sig(k)}^{EUF-CMA}$ is:
		\begin{equation*}
			\begin{cases}
				1, & \text{if } \text{Verify}(pk, m^*, \sigma^*)=\text{True and } m^*\notin \{m_i\}_{i=1}^n;\\
				0, & \text{Otherwise}
			\end{cases}
		\end{equation*}
	\end{algorithmic}
\end{algorithm}

The success probability (or Advantage ($Adv$)) of $\mathcal{A}$ can be defined as:
\begin{equation*}
	Adv(\mathcal{A}_{Sig(k)}^{EUF-CMA}) = Prob[Exp_{Sig(k)}^{EUF-CMA} =1]
\end{equation*}
The signature scheme $sig(k)$ is EUF-CMA secure if $Adv(\mathcal{A}_{Sig(k)}^{EUF-CMA})$ of any probabilistic polynomial time (PPT) adversary $\mathcal{A}$ is negligible concerning the security parameter $k$ \cite{19}.

\begin{theorem}
	Consider $H$ be a cryptographically secure, collision-resistant hash function modeled as a random oracle. The proposed signature scheme is Existentially Unforgeable under Chosen Message Attacks (EUF-CMA), assuming that the Discrete Logarithm with Matrix Decomposition Problem (DL-MDP) is computationally hard.
	
	\begin{proof}
		We prove this theorem by contradiction. Suppose there exists an adversary $\mathcal{A}$ that can win EUF-CMA game with non-negligible probability. We will construct an oracle machine $\mathcal{O}^{\mathcal{A}}$ that uses $\mathcal{A}$ to solve DL-MDP, thereby contradicting the assumed hardness of DL-MDP.
		We proceed through a sequence of games: $G_0$, $G_1$ and $G_2$, where each game introduces a slight change to the previous one. Let  $Prob[G_i]$ denote the  probability that $\mathcal{A}$ wins game $G_i$.
		
		\begin{itemize}
			\item {$\mathbf{G_0}$}: This game represents the standard EUF-CMA security experiment for the proposed signature scheme. Therefore, we have: $$Adv(\mathcal{A}_{sig(k)}^{EUF-CMA})=Prob[Exp_{sig(k)}^{EUF-CMA}] = Prob[G_0]$$.
			
			\item {$\mathbf{G_1}$}: Game $\mathbf{G_1}$ is identical to $\mathbf{G_0}$, except for the following modification: For each $j=1,\dots,n$, whenever  $\mathcal{A}$ queries the hash oracle $H$ on a message $msg_j$, the oracle
			$\mathcal{O}^{\mathcal{A}}$ substitutes the hash outputs with 
			$$m_j = f(\overline{W_x}\cdot (\sigma_0-f(\overline {W_{\theta}}\theta)))\mid\mid f(\overline{W_x}\cdot(\sigma_1-f(\overline{W_{\theta}}\theta)))$$
			where $\sigma_j$ is chosen uniformly at random from $\mathbb{Z}_p^n$. The signature query responses are based on these values.
			Since $H$ is assumed to be a cryptographically secure collision-resistant hash function, the output of $H$ is computationally indistinguishable from a truly random function. Hence, $\mid Prob(G_1)-Prob(G_0)\mid$ were non-negligible, it would imply that $\mathcal{A}$ can distinguish the simulated hash from a truly random function, which contradicts the security of $H$. Therefore, $\mid Prob(G_1)-Prob(G_0)\mid = \epsilon_1(k)$, where $\epsilon_1(k)$ is negligible function.
			
			\item {$\mathbf{G_2}$}: Game $g_2$ is the same as $G_1$, except that when  $\mathcal{O}^{\mathcal{A}}$ queries $H$ on the challenge message  $msg^{*}$, the oracle returns a uniformly random value $m^*\in\mathbb{Z}_p^n$, independent of any actual signature. By a similar indistinguishability argument, we conclude: $\mid Prob[G_2]-Prob[G_1]\mid  = \epsilon_2(k)$ for some negligible function $\epsilon_2(k)$.
		\end{itemize}
		By the triangle inequality: 
		\begin{align*}
			&\mid Prob[G_2] - Prob[Exp_{Sig(k)}^{EUF-CMA} = 1] =\\ 
			&\mid Prob[G_2]-Prob[G_0]\mid \le \mid Prob[G_2]-Prob[G_1]\mid +\\&\mid Prob[G_1]-Prob[G_0]\mid = \epsilon_1(k)+\epsilon_2(k) = \epsilon(k),
		\end{align*}
		where $\epsilon(k)$ is negligible function. Thus we have:
		$
		Prob[G_2] = Prob[Exp_{sig(k)}^{EUF-CMA}=1] \pm \epsilon(k).
		$
		Now, if the probability $Prob[G_2]$ is non-negligible, then $\mathcal{A}$ can produce a forgery $(\sigma_0,\sigma_1)$ for message $msg^*$ such that:
		$$
		f(\overline{W_x}\cdot(\sigma_0^*-f(\overline{W_{\theta}}\cdot\theta)))\mid\mid f(\overline{W_x}\cdot(\sigma_1^*-f(\overline{W_{\theta}}\theta))) = m^*,
		$$
		with non-negligible probability. Therefore, the oracle machine $\mathcal{O}^{\mathcal{A}}$ using $\mathcal{A}$, is capable of solving the DL-MDP problem. This contradicts the assumption that DL-MDP is computationally hard. Hence, our initial assumption that $\mathcal{A}$ can win the EUF-CMA game with non-negligible probability must be false.
		We conclude that 
		$Adv(\mathcal{A}_{sig(k)}^{EUF-CMA})=Prob[Exp_{sig(k)}^{EUF-CMA}=1]$ is negligible. This completes the proof.	
	\end{proof}
\end{theorem}

\begin{lemma}
	Consider $H$ be a cryptographically secure, collision-resistant hash function modeled as a random oracle. The proposed signature scheme is existentially unforgeable under known message attacks, assuming that the Combined Discrete Logarithm with Matrix Decomposition Problem (DL-MDP) is computationally hard.
\end{lemma}

\begin{lemma}
	If an existential forgery exists then the DL-MDP can be solved.
\end{lemma}



\subsection{Analysis of DL-MDP When Synaptic Weight is Non-Diagonalizable}
We prove that the DL-MDP problem remains computationally intractable in polynomial time using matrix eigen decomposition, even  when the synaptic weight matrix 
$W$ is diagonalizable.

\begin{theorem}
	Let the synaptic weight $W\in\mathbb{Z}_p^{2\times 2}$ be a diagonalizable matrix with distinct eigenvalues $\lambda_1,\lambda_2\in\mathbb{Z}_p^*$, and let 
	$P\in\mathbb{Z}_p^{2\times 2}$ be the matrix of eigenvectors such that 
	$W = PDP^{-1}$, where $D = diag(\lambda_1,\lambda_2)$. Given the equation $B = L\cdot PD^aP^{-1}$, recovering the exponent $a\in\mathbb{Z}_p^*$ and the permutation matrix  $L$ requires solving system of nonlinear equations involving discrete logarithms upto permutations $L$ with respect to multiple distinct bases, which is computationally hard.
	\begin{proof}
		Assume $W\in\mathbb{Z}_p^{2\times 2}$ is diagonalizable, i.e., 
		\begin{equation*} W = PDP^{-1},~\text{where}~
			~D = 
			\begin{bmatrix}
				\lambda_1 & 0\\
				0 & \lambda_2 
			\end{bmatrix}~\text{and}~ \lambda_1,\lambda_2\in\mathbb{Z}_p^{*}.
		\end{equation*}
		The equation $B = L\cdot W^a$ transforms into $B = L\cdot PD^aP^{-1}$. Since $L$ is a permutation matrix, there are 2 possible permutations. For each $L$, define: $D^a = P^{-1}BP$ where $D^a = diag(\lambda_1^a,\lambda_2^a)$. Therefore, computing $D^a$ reveals both $\lambda_1^a$ and $\lambda_2^a$, up to permutation. To recover $a$, we must solve: 
		\begin{equation*}
			\lambda_1^a = x\pmod p ~\text{and}~
			\lambda_2^a = y\pmod p
		\end{equation*}
		for known $x,y\in\mathbb{Z}_p$. Since, $\lambda_1\neq \lambda_2$, this constitutes a system of multiple-base Discrete Logarithm Problem, which is known to be computationally hard as it is non-linear in nature over the finite fields. There is no such quantum algorithms exist to solve simultaneous discrete logs with different bases efficiently unless special structure exists.
	\end{proof}
\end{theorem}

Without loss of generality, the above result can be generalized into the broader setting, and formalized as:
\begin{theorem}
	Let the synaptic weight $W\in\mathbb{Z}_p^{n\times n}$ be a diagonalizable matrix with distinct eigenvalues $\lambda_1,\lambda_2,\dots,\lambda_n\in\mathbb{Z}_p^*$. Let 
	$P\in\mathbb{Z}_p^{n\times n}$ be the matrix of corresponding eigenvectors such that 
	$W = PDP^{-1}$, where $D = diag(\lambda_1,\lambda_2,\dots,\lambda_n)$. Given the equation $B = L\cdot PD^aP^{-1}$, recovering the exponent $a\in\mathbb{Z}_p^*$ and the permutation matrix  $L$ requires solving system of nonlinear equations involving discrete logarithms upto permutations $L$ with respect to multiple distinct bases. The presence of the unknown permutation $L$ further obscures the eigenvalue alignment, making the system nontrivial even when diagonalization is known. As such, this inversion task is computationally hard under the standard assumptions that solving the system of non-linear Discrete Logarithm Problem over the finite fields is computationally not feasible even for quantum computers in polynomial time setting.
\end{theorem}

\subsection{Analysis of DL-MDP when synaptic weight is non-diagonalizable}
Suppose the matrix $W\in\mathbb{Z}_p^{n\times n}$ is non-diagonalizable. Then it can be expressed in terms of its Jordan canonical form: $W = PJP^{-1}$, where $J$ is a block-diagonal matrix composed of Jordan blocks $J_i$, such that $J = diag(J_1,J_2,\dots, J_k)$. Each Jordan block  $J_i\in\mathbb{Z}_p^{m_i\times m_i}$ has the form: 
\begin{equation*}
	J_i = \begin{bmatrix}
		\lambda_i & 1 & 0 & \cdots & 0 \\
		0 & \lambda_i & 1 & \cdots & 0 \\
		\vdots & \vdots & \vdots & \ddots & \vdots\\
		0 & 0 & 0 & \cdots \lambda_i & 1\\
		0 & 0 & \cdots & 0 & \lambda_i
	\end{bmatrix}\in\mathbb{Z}_p^{n\times n}
\end{equation*}
Each $J_i$ is an upper triangular matrix with the eigenvalue $\lambda_i$ on the diagonal and ones on the superdiagonal. These matrices are not diagonal and their non-diagonal structure complicates exponentiation. Since $W = PJP^{-1}$, we have $W^a = (PJP^{-1})^a = PJ^aP^{-1}$. To exponentiate a Jordan block $J_i$, we use the binomial expansion involving its nilpotent part $N$, which consists of the superdiagonal ones (with $N^{m_i} = 0$):
\begin{align*}
	J_i^a = \lambda_i^aI+\binom{a}{1}\lambda_i^{a-1}N + \binom{a}{2}\lambda_i^{a-2}N^2+\dots\\+\binom{a}{m_i-1}\lambda_{i}^{a-m_i+1} N^{m-1}
\end{align*}
Hence, the matrix power $W^a$ involves both powers of eigenvalues and binomial coefficients modulo $p$, yielding a highly nonlinear structure. Now consider the problem of recovering the exponent $a$ (modulo a masking permutation or transformation $L$) from: $B = L\cdot W^a  = L\cdot PJ^aP^{-1}$ becomes equivalent to solving $P^{-1}L^{-1}BP = J^a$. Thus, recovering $a$ from reduces to solving for the exponent in $J^a$, which is not a classical Discrete Logarithm Problem (DLP). Unlike DLPs in abelian groups, this is a non-abelian matrix power problem, with dependencies on both field operations and binomial structures in $\mathbb{Z}_p$.
Such a problem does not admit any known efficient classical or quantum algorithm, including Shor’s algorithm \cite{4}, which applies only to abelian group structures. Therefore, the DLP over non-diagonalizable matrices with Jordan forms constitutes a hard, structured, non-commutative exponentiation problem, resistant to quantum attacks and suitable for post-quantum cryptographic applications.


\subsection{Direct Attack}
In Section \ref{sec:nntomvc}, we demonstrated how the proposed binary-weight neural network with recurrent attention can be mapped to a non-linear multivariate polynomial system, while preserving the NP-hardness of the original formulation. Since the proposed neural network can be viewed as a generalization of the Clipped Hopfield Neural Network \cite{3}, a similar structure applies to its encoding scheme: the number of attractors selected to represent plain texts is
$P$, resulting in 
$P!$ possible coding matrices. Each coding matrix corresponds to a unique mapping of plaintext to network attractors. 
The key space is defined as: $|\mathcal{K}| = (L_x,L_{\theta},W, A_x,A_\theta,a,Hash)$, where $L_x$ and $L_{\theta}$ are $n\times n$ permutation matrices, each contributing $n!$ possibilities. The total contribution from both is $(n!)^2$. Vectors $A_x$ and $A_{\theta}$ are of length $n$ contributing $n^{2n}$ to the key space. The element $a\in\mathbb{Z}_p\backslash\{0,1\}$, which has  $p-2$ elements. Consequently, the total size of key space is  $\mathcal{K} = (n!)^2\cdot n^{2n}\cdot(p-2)$ which is proportional to $2^{\Theta(n\log n)}$. A brute force attack would require searching a key space of size $(n!)^2\cdot n^{2n}\cdot(p-2)$, which corresponds to a computational complexity of $2^{\Theta(3n\log n+\log p)}$, which is exponential in $n$ and logarithmic in $p$, making the attack computation infeasible for suitably large parameters. For typical cryptographic parameters $n=128$ and $p\approx 2^{128}-159$ a  256-bit prime, the total key space is approximately $2^{3840}$ for classical setting and if a quantum attacker using Grover's algorithm \cite{5} tries to find the key would still requires around $2^{1920}$ operations. This remains computationally infeasible with current and foreseeable quantum technology, ensuring a strong post-quantum security for the chosen parameters.

\subsection{Security Analysis of Threshold Vector}

The security of the synchronization of the shared vector relies on the secrecy of the vector $r\in\mathbb{Z}_p^n$, which in turn is protected by the computational hardness of the Discrete Logarithm Problem (DLP) and the one-way trapdoor property of the employed hash function. An adversary attempting a brute force attack to recover the shared vector $r$ would proceed by iterating over all possible exponents $a\in\mathbb{Z}_p$, computing $W^a$, applying the hash function to $W^a$, and computing the result with the target vector $r\in\mathbb{Z}_p^n$. Assume that computing $W^a$ requires $\mathcal{O}(n^3)$ operations, and that applying the hash function incurs an additional $\mathcal{O}(hash)$ cost per trial. Since there are $p$ possible values of $a$, the overall time complexity of this brute force attack is: $\mathcal{O}(p\cdot(n^3+hash))$.

\section{Operating Characteristics of Signature Scheme}\label{sec:operationchar}
In this section, we analyze the operational characteristics of the proposed signature scheme. 

\begin{theorem}
The number of operations required in KeyGen are at most $\mathcal{O}( n^3\log p)$.
\begin{proof}
	For public key generation, we compute $\overline{W_x} = L_x \cdot W_x^a$ and $\overline{W_{\theta}} = L_{\theta} \cdot W_{\theta}^b$, where $W_x, W_{\theta} \in GL(n, \mathbb{Z}_p)$ and $a, b \in \mathbb{Z}_p^*$. The exponentiation $W^a$ requires approximately $n^3 \log p$ bit operations. Therefore, computing both $W_x^a$ and $W_{\theta}^b$ requires around $2n^3 \log p$ bit operations. Multiplying a permutation matrix $L$ with $W^a$ takes $\mathcal{O}(n^2 \log p)$ bit operations, so computing both $\overline{W_x}$ and $\overline{W{\theta}}$ requires $\mathcal{O}(2n^2 \log p)$ operations. Hence, the total number of bit operations for public key generation is $\mathcal{O}(2n^3 \log p + 2n^2 \log p)$, which is asymptotically proportional to $\mathcal{O}(n^3 \log p)$.
\end{proof}
\end{theorem}

\begin{theorem}
The number of operations required in the Sign algorithm is at most $\mathcal{O}(n^3)$.
\begin{proof}
	Let $M \in \mathbb{Z}_p^n$ be the message to be signed. The signature $\sigma = (\sigma_0, \sigma_1)$ is computed based on $M$ and using Algorithm \ref{algo:sign}. The total number of field operations required to compute $\sigma$ is $\frac{2}{3}n^3+6n^2 - n$, which is asymptotically bounded by $\mathcal{O}(n^3)$.
\end{proof}
\end{theorem}

\begin{theorem}
The number operations required to verify the signature of corresponding given message $M$ is $\mathcal{O}(n^2)$.
\begin{proof}
	To verify whether the signature $\sigma = (\sigma_0,\sigma_1)$ is valid or not. The total number of field operations required is $3n^2 - n$, which is asymptotically bounded by $\mathcal{O}(n^2)$.
\end{proof}
\end{theorem}

Public key cryptosystems based on multivariate polynomials have been introduced in the literature \cite{6}. Solving systems of multivariate polynomials over finite fields commonly known as the Multivariate Quadratic (MQ) problem, which is a well-established NP-hard problem. Traditional approaches to designing such cryptosystems typically involve constructing a central quadratic trapdoor polynomial function and concealing it using affine transformations. This results in a set of multivariate polynomial equations that form the public key.
In contrast, the approach presented in this work takes a different direction. Rather than explicitly constructing multivariate polynomial functions, we employ a neural network framework to generate the necessary matrices. These matrices are then used, following the iterative procedure described in Section \ref{sec:nntomvc}, to produce a digital signature for a given message. The larger key size of any multivariate polynomial based public key signature scheme is the major drawback. In our proposed scheme we try to mitigate this issue by using weights of Neural Network preserving the same security over finite field $\mathbb{Z}_p$. 


The public key size for the proposed signature scheme is mentioned in Section \ref{sec:keygen} is $\overline{W_x}$ and $\overline{W_{\theta}}$ is:
\begin{equation}
\text{Size of public key}:= \lceil \frac{2n^2\cdot \log_2 p}{8}\rceil.
\end{equation}
Similarly, the private key tuple $(L_x,L_{\theta},a,b, \rho, W,A_j)$ requires 
$\lceil\frac{2n\cdot\log_2 n}{8}\rceil + \lceil\frac{3\log_2 p}{8}\rceil +\lceil \frac{n^2\log_2p}{8}\rceil+\lceil \frac{jn\log_2p}{8}\rceil$ bytes of storage, where $\lceil\cdot\rceil$ denotes the ceiling function. Additionally the signature size for given message $M$ according to Algorithm \ref{algo:sign} is $2n^2+7n$ which is proportional to $\mathcal{O}(n^2)$. In Table \ref{table:comparison}, we compare these operational characteristics with the different multivariate polynomial based signature scheme with different security levels.

\begin{table*}[t]
\centering
\caption{Comparison of Signature Schemes at Various Security Levels (Note: The parameters for Proposed Scheme is $(p,n,j)$).}
\label{table:comparison}
\resizebox{\textwidth}{!}{%
	\begin{tabular}{|l|l|l|l|l|l|}
		\hline
		\textbf{Security Level ($\ell$)} & \textbf{Signature Scheme} & \textbf{Hash Length} & \textbf{Signature Size} & \textbf{Public Key Size} & \textbf{Secret Key Size} \\ \hline
		
		\multirow{4}{*}{80} 
		& UOV \cite{15} ($2^8$, 28, 56)         & 224         & 672 bit        & 99.9 kB         & 93.5 kB         \\ \cline{2-6} 
		& Rainbow \cite{20} ($2^8$, 17, 13, 13) & 208         & 344 bit        & 25.1 kB         & 19.1 kB         \\ \cline{2-6} 
		& MQDSS \cite{21} ($2^8$, 26, 26, 81)   & 208         & 8.8 kB         & 58 Bytes        & 64 Bytes        \\ \cline{2-6} 
		& {\bf Proposed Scheme (257, 26, 10)}         & 208         & 1764 bits      & 1570 Bytes      & 1104 Bytes      \\ \hline
		
		\multirow{4}{*}{100} 
		& UOV \cite{15} ($2^8$, 35, 70)         & 280         & 840 bits       & 193.8 kB        & 179.5 kB        \\ \cline{2-6} 
		& Rainbow \cite{20} ($2^8$, 26, 16, 17) & 264         & 472 bits       & 59 kB           & 45 kB           \\ \cline{2-6} 
		& MQDSS \cite{21} ($2^8$, 33, 33, 101)  & 264         & 13 kB          & 65 Bytes        & 64 Bytes        \\ \cline{2-6} 
		& {\bf Proposed Scheme  (257, 33, 10)}        & 264         & 2409 bits      & 2180 Bytes      & 1467 Bytes      \\ \hline
		
		\multirow{4}{*}{128} 
		& UOV \cite{15} ($2^8$, 45, 90)         & 360         & 1080 bits      & 409.4 kB        & 375.9 kB        \\ \cline{2-6} 
		& Rainbow \cite{20} ($2^8$, 36, 21, 22) & 344         & 632 bits       & 136.1 kB        & 102.5 kB        \\ \cline{2-6} 
		& MQDSS \cite{21} ($2^8$, 43, 43, 129)  & 344         & 20.3 kB        & 75 bits         & 64 bits         \\ \cline{2-6} 
		& {\bf Proposed Scheme (257, 43, 10)}         & 344         & 3999 bits      & 3701 Bytes      & 2345 Bytes      \\ \hline
	\end{tabular}%
}
\end{table*}
\section{Conclusion}\label{sec:conclusion}
In this work, we have extended the traditional Hopfield Neural Network (HNN) by introducing binary weights and a recurrent random attention mechanism. This enhancement incorporates structured randomness via an attention vector and leverages the softmax function to maintain non-linearity and system stability. To underpin the cryptographic strength of our design, we introduced a novel hardness assumption that combines the Discrete Logarithm with a matrix decomposition challenge, along with defining appropriate security parameters that offer 128-bit security against both classical and quantum adversaries.
We also proposed a secure bias vector synchronization algorithm grounded in the hardness of the discrete logarithm problem and the trapdoor properties of specific hash functions. Building on this, we developed a public-private key signature scheme in which key pairs are derived from the neural network's weight matrix and masked using a permutation matrix to ensure privacy and security.

Our analysis demonstrates that integrating neural network structures with cryptographic primitives can yield secure and efficient systems. The proposed signature scheme is proven to be secure against existential unforgeability under chosen message attacks (EUF-CMA) and matrix factorization attacks. Additionally, our complexity analysis confirms that brute-force attacks are computationally infeasible under both classical and quantum settings, given the chosen parameters. We further evaluated the security of the threshold vector and quantified the computational cost of key generation, message signing, and verification, providing a comprehensive view of the scheme’s practicality and robustness.

%







%
\end{document}